\documentclass[aps,pra,twocolumn,showpacs,groupedaddress,amsmath,amssymb]{revtex4}
\usepackage{graphicx}
\usepackage{color}
\usepackage{comment}
\usepackage{bm}
\usepackage{amsmath}
\usepackage{latexsym}

\begin{document}

\title{Environmentally-induced exceptional points in elastodynamics}

\author{V.  Dom\'inguez-Rocha$^1$, Ramathasan Thevamaran$^2$, F. M. Ellis$^1$, and T. Kottos$^1$}
\address{$^1$Department of Physics, Wesleyan University, Middletown, CT-06459, USA}
\address{$^2$Department of Engineering Physics, University of Wisconsin, Madison, WI 53706, USA}

\begin{abstract}
We study the nature of an environment-induced exceptional point in a non-Hermitian pair of coupled mechanical oscillators. The mechanical oscillators are a pair of pillars carved out of a single isotropic elastodynamic medium made of aluminum and consist of carefully controlled differential losses. The inter-oscillator coupling originates exclusively from background modes associated with the ``environment'', that portion of the structure which, if perfectly rigid, would support the oscillators without coupling. We describe the effective interaction in terms of a coupled mode framework where only one nearby environmental mode can qualitatively reproduce changes to the exceptional point characteristics. Our experimental and numerical demonstrations illustrates new directions utilizing environmental mode control for the implementation of exceptional point degeneracies. Potential applications include a new type of non-invasive, differential atomic force microscopy and hypersensitive sensors for the structural integrity of surfaces.
\end{abstract}

\maketitle

%

\section{INTRODUCTION}
\label{sec:intro}
The study of exceptional points (EPs) has revealed a variety of fundamental phenomena and spawned next generation technological developments \cite{FGG17,GMKMRC18,MA19}. Examples range from hypersensitive gyroscopes and bio-sensing \cite{HHWGGCK17,M17,COZWY17,W14} to lasing control \cite{HMHCK14,FWMWZ14,HMHHHCK16} and unidirectional invisibility \cite{L11,P14}. Most studies have been performed in the photonics framework \cite{FGG17,GMKMRC18,MA19}, and only few utilized other areas like electronic circuitry \cite{SLLREK12,AYF17,LCTEK18}, acoustics \cite{SDCCRWZ16,FSA16,AP17} and atomic physics \cite{PCSQWJX16, HH17}. Here we develop a platform where EPs are realized by coupling together, via an elastic plate that emulates a complex environment, two identical elastic resonators with differential loss. In contrast to previous studies, the formation of the EPs, are manipulated via a set of ``environmental" plate modes that indirectly control the coupling between the two resonators. Our work paves the way for exciting applications in micro-electromechanical device engineering, were the hypersensitive nature of EPs can be utilized for monitoring structural integrity of surfaces or for realizing a new family of double-cantilevered atomic force microscopes. 

\section{EXCEPTIONAL POINTS FROM PARITY TIME SYMMETRY}
\label{sec:ptep}
\begin{figure*}
  \includegraphics[width=\hsize,keepaspectratio]{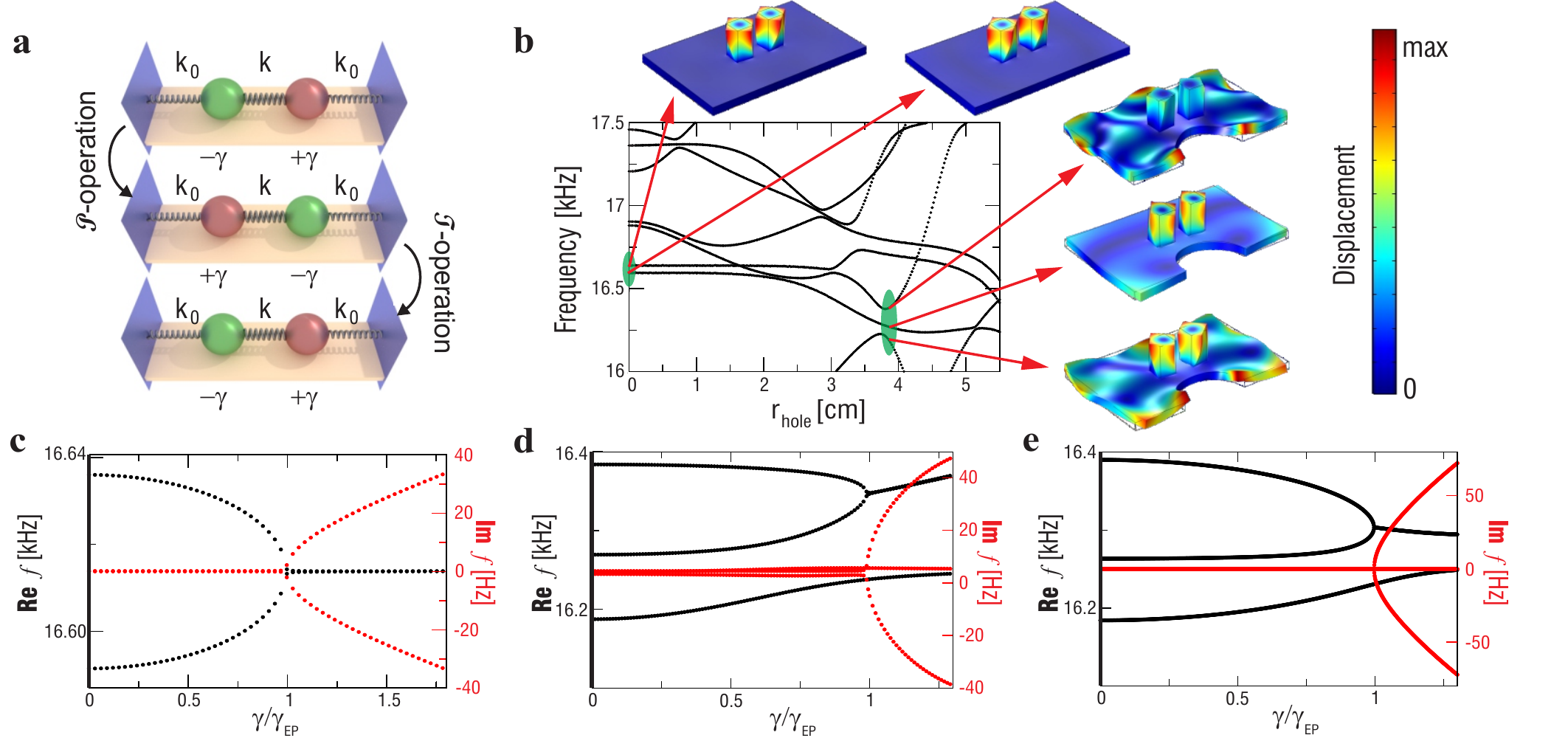}
\caption{(a) A simple ${\cal PT}$-symmetric dimer made by two identical coupled oscillators, with one of them (green) experiencing energy attenuation and the other (red) energy amplification. (b) COMSOL simulations of the parametric evolution of the eigenfrequency spectrum of an elastic dimer system consisting of two pillars formed in close proximity to a base. The two pillars are located symmetrically along the long axis of a $20 cm$ by $12.8 cm$ plate, $1 cm$ thick. Each pillar has a $4 cm$ height with a square cross section $2 cm$ on a side, with their centers displaced by $4 cm$. The varied parameter is the radius 
$r_{\rm hole}$ of a semicircular indent at a position that preserves the mirror symmetry of the structure. The gain/loss parameter $\gamma$ is zero. The set of modes that are investigated in the presence of gain/loss contrast are highlighted with a green ellipse. At the insets we show the numerically evaluated (using COMSOL) corresponding eigenmodes. (c) The real and imaginary parts of the eigenfrequencies for the fundamental torsion modes, corresponding to $r_{\rm hole}=0$, are shown vs. the gain/loss parameter $\gamma$. (d) The same for the three nearby modes associated with $r_{\rm hole}\approx3.8cm$. (e) The same as previously but for the eigenmodes of the effective Hamiltonian Eq. (\ref{3by3}). The dimensionless parameters used are $\kappa=6.14 \times 10^{-3}, \lambda_1=-0.9\times10^{-3}=-\lambda_2, \nu=-2\times 10^{-3}$ while $\omega_0=2\times \pi\times16.290\times 10^3 Hz$. 
}
\label{fig1}
\end{figure*}

Theoretically discussed more than fifty years ago \cite{kato66}, EPs are non-Hermitian degeneracies associated with the coalescence of two eigenvalues, and their corresponding eigenvectors \cite{heiss12}. The simplest example is a parity-time (${\cal PT}$)-symmetric dimer \cite{BB99} consisting of two identical oscillators of mass $m$ and spring constant $k_0$, having resonance frequency $\omega_0 =\sqrt{k_0/m}$. The oscillators are coupled together with a spring with Hooke's constant $k$, see Fig.~\ref{fig1}(a). Each oscillator is equipped with opposing power flow, a gain and loss of strength $\gamma = b/(m \omega_0)$ expressed in terms of an equivalent linear drag and anti-drag forces $f_\mp = \mp bv$ where $v$ is the speed of each oscillator and $b$ the drag coefficient. Such a system is not invariant under time-reversal (${\cal T}$) i.e. inversion of the flow of time $t\rightarrow -t$. It also violates parity symmetry, i.e., it is not invariant under spatial reflections $x \rightarrow -x$. Instead, its equations of motion are invariant under joint ${\cal PT}$-symmetry for all $\gamma$-values.

This can be seen in the equations of motion, which in the frequency domain ($e^{i \omega t}$) are given by
\begin{equation}
\begin{pmatrix}
1 + \kappa - iu\gamma - u^2 & -\kappa \\
-\kappa & 1 + \kappa + iu\gamma - u^2
\end{pmatrix}
\begin{pmatrix}
x_1 \\
x_2
\end{pmatrix}
=0
\label{mpt_dimer}
\end{equation}
where $u=\omega/\omega_0$ is the frequency relative to the isolated oscillator frequency, and $\kappa = k/k_0$ measures the strength of the inter-oscillator coupling spring. The loss (drag) in Eq.(\ref{mpt_dimer}) is applied to particle 1 and the gain (an anti-drag) is applied to particle 2, evident by the signs of the respective terms involving the gain/loss parameter $\gamma$ defined earlier. The ${\cal PT}$-symmetry is evident from the parity interchange operation (switch the rows and columns) and the time reversal operation (change the sign of $i$) which together return to the same relations.

We make a connection to coupled mode theory (CMT) by casting the equations of motion for this ${\cal PT}$-dimer into a Liouvillian-like eigenvalue form 
\begin{equation}
\Omega \psi = \omega \psi,\quad
\Omega = \omega_0 
\begin{pmatrix}
1 +{\kappa - i \gamma\over 2}  & -{\kappa \over 2} \\
-{\kappa \over 2} & 1 +{\kappa + i \gamma \over 2}
\end{pmatrix}
\label{std_pt}
\end{equation}
where $\psi$ is a vector of oscillator amplitudes. This form assumes weak coupling so that $u \approx 1$ and greatly simplifies the analytic description, with the factor of two originating from the epansion of $1-u^2$. (see the Appendix)

The eigenfrequencies and eigenmodes of this system are 
\begin{multline}
\hspace{5mm}
\omega_\pm=1+{\kappa\over 2}\mp {1\over 2}
\sqrt{\kappa^2-\gamma^2} 
\\
\text{and }\hspace{5 mm} \psi_\pm=\left(\frac{i\gamma\pm \sqrt{\kappa^2-\gamma^2}}{\kappa}, 1\right)^T
\label{d_loss}
\end{multline}
respectively. In the {\it exact phase} $\gamma<\kappa\equiv\gamma_{\rm EP}$, the coupling between the spatially separated gain and loss elements is capable of exactly communicating a balanced flow of energy: the modes have real eigenvalues and identical, steady-state oscillatory magnitudes. In this phase the eigenvectors $\psi_{1,2}$ are also eigenvectors of the ${\cal PT}$-operator. In the 
{\it broken phase} $\gamma>\gamma_{\rm EP}$, the flow of energy overwhelms the coupling and effectively decouples the two oscillators into modes that are ${\cal PT}$ images, $\psi_2={\cal PT}\psi_1$, with a single real frequency and conjugate imaginary parts. The $\psi_2 (\psi_1)$ mode, describing displacements predominantly on the gain (loss) side, grow (decay) in time. The exact and broken ${\cal PT}$-symmetry phases are separated by a sharp transition -- the exceptional point (EP) -- where the eigenvectors $\psi_{1,2}$ coalesce and the frequencies merge with a characteristic square-root singularity $\Delta\omega \sim \sqrt{\gamma_{\rm EP}-\gamma}$. The scenario is generic and applies to other anti-linear symmetries ${\cal LT}$, where ${\cal L}$ is not necessarily the parity ${\cal P}$ but any other linear operator such as mirror symmetry ${\cal M}$ or rotation symmetry ${\cal R}$ \cite{BB99,B07,M11}.

\section{ENVIRONMENTAL MODES}
\label{sec:env}
In our study, each oscillator of the dimer is taken as the fundamental quarter-wave torsion resonance of an elastic pillar with one free end and one fixed end, see insets of Fig.~\ref{fig1}(b). The torsion pillars are formed in close proximity on a base composed of the same elastodynamics medium (aluminum). It has to be stressed that the complexity of the actual system is far larger than the simple coupled oscillators discussed above. Even the one-dimensional approximation of elastic pillars at long wavelengths allows for three distinct modes: bending, compression, and torsion. In three-dimensions, any free surface invokes hybridized bulk modes \cite{graff91}. Here we focus our interest on the lowest torsional pillar modes to avoid the most obvious couplings to out-of-plane plate modes.

When two similar pillars are brought in the proximity of one another, communication of displacement fields through the not-perfectly-rigid base couples the two resonators, lifting their degeneracy. The frequency splitting is an indication of the coupling strength, $\kappa\approx\Delta f\approx 45Hz$ (see green highlight at the left side of Fig.~\ref{fig1}(b). The emerging pair is associated with a hard (out of phase) and soft (in phase) torsional supermodes respecting a ${\cal MT}$ symmetry of the total structure, seen highlighted in green on the left axis of Fig.~\ref{fig1}(b).

Unlike the simple coupled oscillator model, the pillar-pillar coupling involves {\it only} interactions with the modes of the base (environmental modes) leading to far more reaching consequences. In our simulations, we purposely manipulate the influence of these environmental levels to a dramatic degree by introducing a semicircular indent of radius $r_{\rm hole}$ at a position that preserves the mirror symmetry of the structure. We note that such configurations (known as Sinai billiards) lead to chaotic dynamics in the classical (particle) limit with direct consequences to the levels and eigenmodes of the wave system \cite{levels}. A typical signature of chaoticity in the wave mechanics framework is the formation of avoided crossings between nearby levels \cite{levels} (level repulsion) as $r_{\rm hole}$ increases, see Fig.~\ref{fig1}(b). 

We use COMSOL Multiphysics\cite{comsol} to investigate of the system as a function of the indent size, and inspect the behavior at key values of $r_{\rm hole}$ as a function of a balanced gain/loss contrast $\gamma$ in the pillars by imposing opposite imaginary parts to their shear modulus i.e. $G=G_0 (1 \pm i\gamma)$ where $G_0=25 GPa$ is the shear modulus for aluminum. In the case of a simple two-level interaction (e.g. at $r_{\rm hole}=0$ -- see left level pair in Fig.~\ref{fig1}(b) -- the normal modes approach one another as $\gamma$ increases, and eventually coalesce at some $\gamma_{\rm EP}$ which is determined by the coupling strength $\kappa$ between the two torsional modes, see Fig.~\ref{fig1}(c). The symmetry-violation scenario is the one common to the ideal ${\cal PT}$-oscillator model, Fig.~\ref{fig1}(c) as well as the coupled mode theory model of Eq.~\ref{std_pt}.

\begin{figure*}
  \includegraphics[width=\hsize,keepaspectratio]{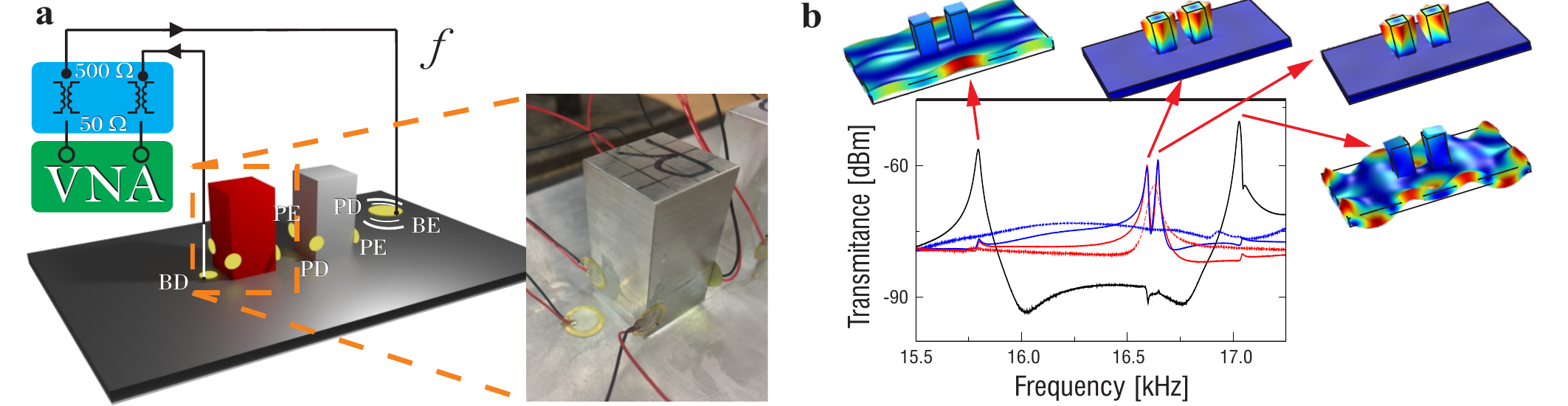}
\caption
{(a) Photograph of the experimental set-up. The pillar dimer with its plate environment is machined from a single piece of 6061 aluminum. The two pillars are located symmetrically along the long axis of a $20 cm$ by $12.5 cm$ plate, $1 cm$ thick. Each pillar has a $4 cm$ height with a square cross section $2 cm$ on a side, with their centers displaced by $4 cm$. Weakly-coupled piezoelectric transducers allow elastic-wave transmission to be measured by a Vector Network Analyzer (VNA in the figure) through transformers. Plate-to-plate, or pillar-to-pillar transmission can be chosen to explore various modes with differing sensitivity. The lower right inset shows a detail of transducers placed to couple predominantly to the torsion modes. (b) Transmission spectrum of the two pillars structure for different transducer configurations. Of the many modes present, this frequency range spans the two coupled torsional modes at approximately 16.6 kHz and the two closest plate modes on either side. The relative surface displacements are also shown with violet to red spanning zero to a linear maximum. The black solid line shows the base-to-base transmission and the blue and red lines show the inter- and intra-pillar transmissions respectively. The symmetry of the nearest modes illustrates that the inter-pillar coupling must be mediated by other modes, further away in frequency. 
}
\label{fig2}
\end{figure*}

While the two-level interaction physics is typically captured by the phenomenological system of Eq. (\ref{std_pt}), a three-level (or more) interaction can capture features inherent to an environment without any obvious analogue in the coupled oscillator system. We have identified in our simulations such a three-level interaction scenario -- see the encircled three levels at $r_{\rm hole} \approx 3.8cm$ in Fig.~\ref{fig1}(b). In this case, the gain/loss contrast couples the upper level pair having frequency separation (at $\gamma=0$) $\Delta f\approx 115Hz$ larger than $\Delta f\approx 88Hz$, associated with the lower two levels, see Fig.~\ref{fig1}(d). The three-level EP-scenario can be modeled using a $3\times 3$ CMT Hamiltonian
\begin{equation}
\Omega \psi = \omega \psi,\quad 
\Omega =\omega_0
\begin{pmatrix}
1+i\gamma  & -\kappa  & \lambda_1\\
-\kappa  & 1 - i \gamma&\lambda_2\\
\lambda_1  & \lambda_2&1+\nu\\
\end{pmatrix}
\label{3by3}
\end{equation}
where $\kappa$ describes generic coupling, associated with a background sea of spectrally distant base modes between the two upper levels. The third level, having relative frequency detuning $\nu$ (relative to $\omega_0$), describes a particular environmental mode which has notably significant spatial {\it and spectral} overlap with the two torsional modes. In the most general case, each of the two torsional levels interact in a different manner with the environmental level. These interactions are described by the coupling constants $\lambda_{1,2}$ and their influence in the three-level EP formation is demonstrated in Fig.~\ref{fig1}(e). For an appropriate choice of the CMT parameters, most notably $|\lambda_2|=|\lambda_1|$, we observe an EP formation which is qualitatively the same as the one found in Fig.~\ref{fig1}(d). 

We also note that either $\kappa$ or $\lambda$ alone ensure an exceptional point: we include both to accommodate a specific interaction strength for the interfering mode ($\lambda$) distinct from the environmental sea ($\kappa$) of other modes. Their values allow independent control of the relative separation of three modes away from the exceptional point. The upward skewing of the COMSOL result near the exceptional point cannot be captured by the three-level model with constant coupling terms.

\section{EXPERIMENTAL EXEPTIONAL POINT}
\label{sec:eep}

The implementation of non-Hermitian gain and loss mechanisms for the realization of EPs can be achieved via piezo-electric elements attached to each of the pillars. An alternative, experimentally simpler approach, is to introduce controllable differential loss externally applied to one of the pillars. Its downside, however, is that such differential loss configurations require a greater attention in the design of the structure. Specifically, a weak overlap of the environmental modes with the torsional modes of the pillars has to be engineered, thus enforcing the {\it weak} pillar-pillar coupling regime necessary for the realization of EP singularities. In fact, it can be shown (see the Appendix) that in this limit, the CMT model has essentially the same form as Eq.~\ref{std_pt} with the loss terms unequal and both positive:
\begin{equation}
\Omega \psi = \omega \psi,\quad
\Omega = \omega_0 
\begin{pmatrix}
1 +{\kappa + i \gamma_d\over 2}  & -{\kappa \over 2} \\
-{\kappa \over 2} & 1 +{\kappa + i \Gamma \over 2}
\end{pmatrix}.
\label{difep}
\end{equation}
In the symmetric phase, the mean loss is shared equally by both modes (identical imaginary parts, rather than zero) whereas the broken phase modes essentially decay according to their individual damping, with $\Gamma$ the externally imposed damping and $\gamma_d$ the intrinsic damping.

Having designed the experimental pillar/base system (in the absence of the indent, see Fig.~\ref{fig2}(a) to assure the {\it weak coupling limit}, small piezoelectric transducers are carefully attached (see Methods) to various positions in order to study the transmission spectrum through different paths. A network analyzer excites one of the transducers and receives the signal transmitted to a second transducer, both through audio transformers to improve the transducer impedance matching to the analyzer.

Fig.~\ref{fig2}(b) illustrates the transmission spectrum through various paths. Note that the curves in the figure indicate that all modes are only weakly hybridized, in contrast to the modes in Fig.~\ref{fig1}(b) with a large cut-out. The torsion mode doublet is observed in the plate transmission, and the plate modes are observed in the pillar-pillar torsion transmission, while their oscillator strengths are appropriately emphasized. The figure encompasses the torsion doublet ($\approx16.6 kHz$) along with the closest plate modes (15.6 kHz and 17.0 kHz) on either side, accompanied by insets showing their surface displacements. These plate modes correspond to the closest modes of Fig.~\ref{fig1}(b) at $r_{hole}=0$, the lower connecting to the mode seen rising into the graph at $r_{hole} \approx 3.8$.

\begin{figure}
\includegraphics[width=8 cm]{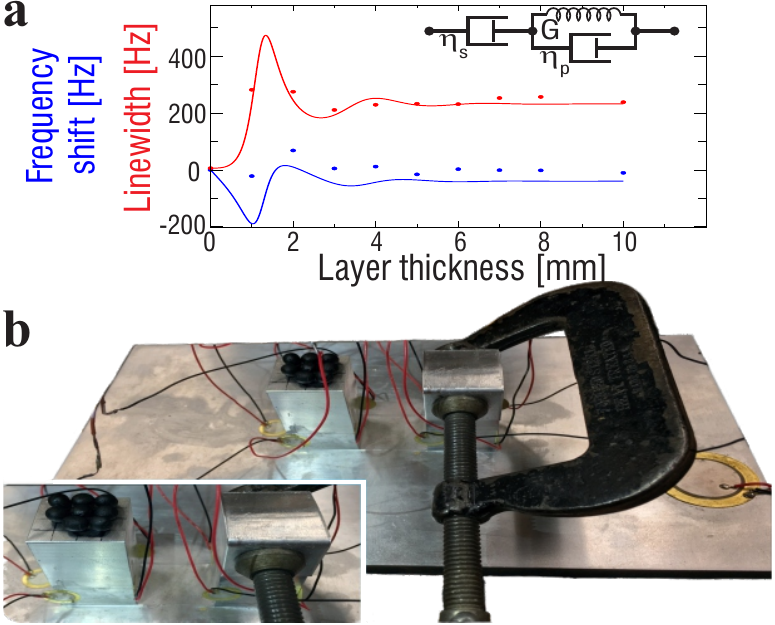}
\caption
{
(a) Experimental measurement of the frequency shift and linewidth resulting from varying thicknesses of putty applied uniformly over the whole top surface of a pillar. This illustrates that for thicknesses over approximately $2mm$, the the damping becomes independent of the thickness and the frequency shift due to mass loading is effectively zero due to the nature of the complete boundary layer. The solid lines are fits to the simple linear viscoelastic model shown in the inset. (b) Image of the putty application creating the incremental loss for the exceptional point data presented in Fig.~\ref{fig4}. The imposed damping was directly measured and the frequency shift was confirmed to be proportionally consistent with the results for the thick uniform layers.
}
\label{fig3}
\end{figure}

The differential loss is physically introduced by small putty balls pressed into the top of one of the pillars. At 16 kHz, materials can interact with a shearing motion of the substrate (the pillar top) through a fairly thin boundary layer that adds dissipation proportional to the area of contact without a significant mass loading. Fig.~\ref{fig3}(a) shows experimental confirmation of this phenomenon, where the single pillar resonance was measured as the complete top of one the pillars was subjected to a uniform layer of increasing thickness with the other pillar disabled. The points show the measurements at different layer thicknesses and the lines show the fit to the model shown in the inset. In blue is the frequency shift experienced by the pillar, while the red shows the dissipative linewidth in the same units. Thickness dependence occurs only when the boundary layer is not contained within the putty layer, and the observed lack of mass loading under these circumstances requires that the boundary layer also not be heavily over-damped so that out-of-phase motion in the complete boundary layer nearly cancels the mass loading. Details of the model are included in the Appendix.

\begin{figure}
\includegraphics[width=8 cm]{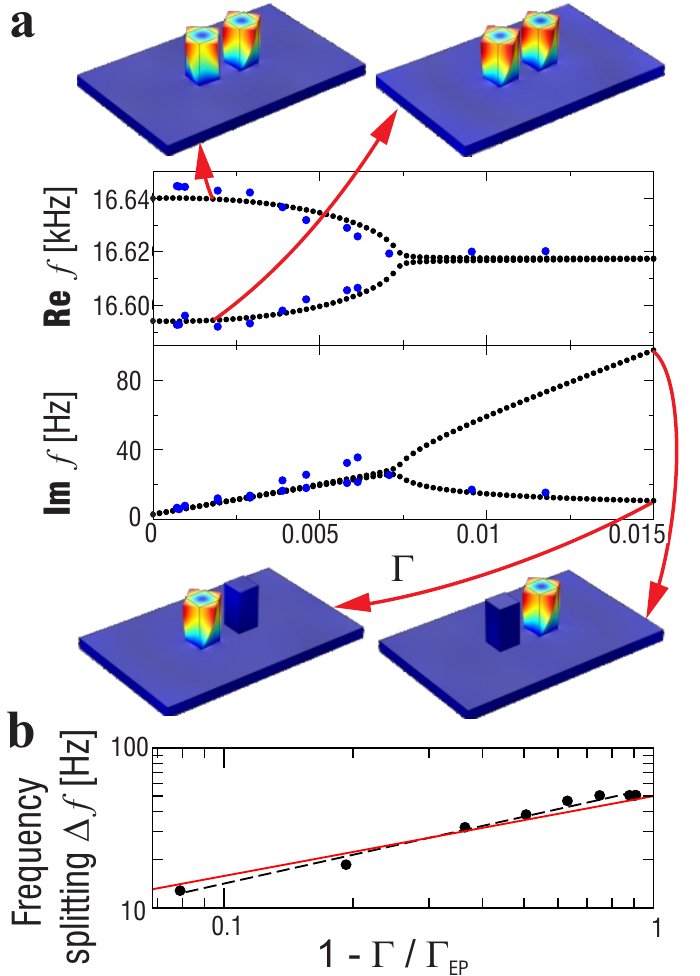}
\caption
{(a)Numerical calculations (black dots) and experimental data (blue dots) of the real and imaginary parts of the eigenfrequencies of the two pillars oscillating in their first torsional modes as a function of the loss parameter $\Gamma$. The experimental results 
are in a good agreement with the numerical simulation. In the same subfigure we report the modes of the system for two typical cases of the differential loss $\Gamma$. (b) Log-log plot of the measured frequency splitting $\Delta f$ versus $1-\Gamma/\Gamma_{\rm EP}$ near the exceptional point. The red solid line shows the square-root law $\delta f\sim \sqrt{1-\Gamma/\Gamma_{\rm EP}}$ while the black dashed line is the best linear fit $\left(1-\Gamma/\Gamma_{\rm EP}\right)^{\alpha}$ with $\alpha\approx 0.6$.
}
\label{fig4}
\end{figure}

With one pillar fixed at its weakly damped intrinsic aluminum loss $\gamma_d$, the additional differential loss was varied by placing small (though larger than the boundary layer) putty balls incrementally to the top of the other pillar. For each increment, the damping was first calibrated by direct measurement of the loss factor $\Gamma$ of the damped pillar with the other pillar's resonance temporarily moved out of the picture by a clamp, see Fig.~\ref{fig3}(b). With the clamp removed, the experimental transmission spectra from the weakly damped side to the damped side was measured, and fit to the function 
 \begin{multline}
t(u)=t_0+ \\
\frac{\gamma_d A e^{i\phi}}{(1-u^2+\kappa+iu\gamma_d)-\kappa^2/(1-u^2+\kappa+\epsilon+iu\Gamma)}
\label{transmit}
\end{multline}
describing the expected for an ideal pair of coupled oscillators, driven on the weakly damped side and sensed on the differential loss side. The parameter $\epsilon$ was included as an additional confirmation of the small size of the mass loading. The eigenfrequencies, extracted from the equivalent un-driven modes using the same parameters are reported with blue dots in Fig.~\ref{fig4}(a). In the same figure we also compare with the numerically calculated eigenmodes of this pillar system using COMSOL simulations. In these simulations we have used as only other input the independently determined experimental loss $\Gamma$ for the loss-pillar and one overall frequency shift adjusted via the exact choice of shear modulus. Poisson's ratio was previously fixed to best match the inherently damped pillar and plate modes of Fig.~\ref{fig2}. The data after the EP is unable to significantly resolve the linewidth of the broader resonances due to their merger into the background of the other modes. A further analysis of the eigenfrequency behavior around the EP is shown in Fig.~\ref{fig4}(b) where we plot the eigenfrequency difference versus the loss parameter $\Gamma$. By considering the logarithmic behavior of this curve, we find that the frequency splitting $\Delta f$ near the exceptional point scales as $\Delta f\sim \sqrt{1-\Gamma/\Gamma_{\rm EP}}$, thus confirming the existence of an EP singularity at a critical value $\Gamma_{\rm EP}$ in the case of {\it weak coupling}.

\section{CONCLUSIONS}
\label{sec:conc}
In this work we have investigated an aluminum double-torsion-pillar system that demonstrates an EP behavior in the elastodynamic regime for a system where the inter-oscillator coupling is completely mediated by environmental modes. We show that in the case of differential losses, a necessary condition for the emergence of EPs is a weak pillar-pillar base-mediated coupling. Our work provides a new versatile platform for the study of non-Hermitian wave physics and opens up new possibilities in engineering EP singularities through judicious manipulation of background environmental modes.

\section*{ACKNOWLEDGEMENTS}
\label{sec:ackn}
We acknowledge partial support from an NSF CMMI-1925530;1925543 grants (RT, FME, TK) and from ONR N00014-16-1-2803 and AFOSR via MURI Grant No. FA9550-14-1-0037 (VDR, TK).

\section*{APPENDIX}
\label{sec:append}
{\bf Coupled mode theory of classical mechanical oscillators:} If the coupling is weak, both $\kappa$ and $\gamma$ can be taken as the same order, acknowledging that the range of interest for $\gamma$ will be limited by $\kappa$. Expressing the frequency $u$ in the matrix of Eq.~\ref{mpt_dimer} as $u=1+\Delta$, with $\Delta$ also of the same order as $\kappa$, the $1-u^2$ and $\gamma (1+\Delta)$, to first order in $\Delta$, are replaced by $-2\Delta$ and $\gamma$ respectively. The standard eigenvalue relation for $\Delta$, after dividing out $2$, is 
\begin{equation}
\bm{U} \vec{x} = \Delta \vec{x};\quad{\rm where}\quad
{\bf U}=\begin{pmatrix}
{\kappa - i\gamma\over 2}& -{\kappa\over 2} \\
-{\kappa\over 2} & {\kappa + i\gamma\over 2}
\end{pmatrix}
\label{U_dimer}
\end{equation}
Eq.~\ref{std_pt} is then expressing a generic tight binding eigenvalue relation with $\bm{\Omega} = \omega_0 (\bm{U}+\bm{I})$ directly in terms of the original simple harmonic oscillator parameters.

The above expressions are for a balanced gain and loss characterized by the single gain/loss parameter $\gamma$. This is the $\cal{PT}$-symmetric case which has and exceptional point separating it's eigenmode spectrum into an exact phase, with two real frequencies having zero imaginary parts, and a broken phase having one real frequency shared by two conjugate imaginary parts \cite{GMKMRC18}. If this gain/loss balance is relaxed, the system is no longer $\cal{PT}$-symmetric. Including an overall loss term, $\gamma_0$ in Eq.(\ref{mpt_dimer}) expresses an unbalanced loss in an explicit differential form. Eq.~\ref{mpt_dimer} becomes
\begin{multline}
\hspace{-5mm}
\begin{pmatrix}
1 + \kappa + iu(\gamma_0-\gamma) - u^2 & -\kappa \\
-\kappa & 1 + \kappa + iu(\gamma_0+\gamma) - u^2
\end{pmatrix}
\begin{pmatrix}
x_1 \\
x_2
\end{pmatrix} \\
=0
\label{d_loss}
\end{multline}
where the eigenvalues $u$, found as the roots of the associated secular equation, demonstrate a strictly singular 
exceptional point only when $\gamma_0 = 0$. Again, however, in the weak coupling limit, the exceptional point signature is restored. The diagonal presence of the $i \gamma_0$ term, in the lowest order matrix form without the $u$, allows a simple shift of the eigenfrequencies by $i \gamma_0 /2$. Though not $\cal{PT}$, this $2 \times 2$ coupled mode theory form does retain a strict exceptional point. 

We have heuristically determined the eigenvalues of the ideal damped mass-and-spring system expressed by the matrix of Eq.~(\ref{d_loss}) in the limit of small $\kappa$ with one oscillator having a fixed small loss, $\gamma_d = \gamma_0-\gamma$, while the other oscillator's imposed loss, $\Gamma=\gamma_0+\gamma$, is increased. This is a better match to the actual experimental situation where the damping factor imposed on the loss pillar is significantly larger than that of the background aluminum, as illustrated by the small intercept for the imaginary part of the frequency shown in Fig.~\ref{fig4} of the main text. For this case, we find that the exceptional point {\it region} (vs. $\Gamma$, as in Fig.~\ref{fig4}(a) for this small-loss/high-loss mechanical system is centered at
\begin{eqnarray}
\Gamma_{\rm EP} = 2(\sqrt{1+2\kappa}-1)+\gamma_d \label{eq:epg} \\
\frac{\omega_{\rm EP}}{\omega_0} = \frac{\sqrt{1+2\kappa}(1+i)+1-{\gamma_d}^2 + i(\gamma_d-1)}{2} \label{eq:epf}
\end{eqnarray}
with the eigenmode splitting at the above position being
\begin{equation}
\frac{\Delta \omega}{\omega_0}  = \frac{2}{3}\kappa^{3/2}.
\label{eq:eps}
\end{equation}
These relationships provide guidance to how the exact simple harmonic mechanical system begins to deviate from the weak coupling limit as the coupling strength increases.

{\bf Interrogation of the Elastodynamics Modes:} In total, ten bending-mode transducers were attached to the system. Plate modes are more efficiently observed by transmission through a pair of transducers attached flush to surface of the plate, with the bending stresses coupling to all of the fundamental forms of bulk modes – one compression and two shear polarizations –  due to their hybridization into free surface modes of the plate. Alternatively, since the torsion modes were the primary focus of this study, particular care was taken to maximize the pillar-pillar torsion transmission while minimally influencing the mode symmetry. To this end, the piezoelectric elements were attached in the chirally symmetric fashion seen in Fig.~\ref{fig2} of the main text, with a small overhang at the pillar edges. This overhang translating the piezo bending motions predominantly into torques applied to the pillars, while canceling the direct pillar bending coupling created by their non-overhanging portions.

Each pillar has two pairs of transducers allowing transmission through a single pillar, or from one pillar to the other, with each opposing pair used in parallel. These transmission path options are used to confirm the equal participation of both pillars in the normal modes, as seen by the close matching of the single pillar doublet with that of the pillar-pillar doublet in Fig.~\ref{fig2}(b). The transmission path including the transducers and aluminum structure was connected to the $50 \Omega$ ports of a Keysight E8050A network analyzer with each port converted to approximately $500 \Omega$ with a United Transformer Type LX-30S audio transformer. In spite of this, even on resonance, the overall transmission is still relatively weak, $\approx -60 dB$. This is primarily due to mechanical impedance mismatch of the small, unobtrusive size chosen for the piezo transducers.

{\bf Theoretical Analysis of the Putty Frequency Shift and Dissipation:} The experimental data for the frequency shift and dissipation induced by the application of putty balls to the top of the torsion pillar showed a remarkably simple behavior: essentially no frequency shift and a dissipation proportional to the contact area. This can be understood in terms of a boundary layer interaction with the surface of the free torsion pillar in the context of a combined Maxwell (series) and Voight (parallel) viscoelastic model~\cite{basicVE} illustrated schematically in the inset of Fig.~\ref{fig3}.

The viscoelastic model is applied in three steps, acknowledging the large mechanical impedance difference that allows for a boundary-layer solution: (1) solve for the surface-stress in the linearized, pure-shear, plane-wave viscoelastic medium of arbitrary thickness; (2) express the result as an effective surface mass density and surface drag constant; and (3) solve for the pillar torsion-mode frequency bounded by the derived surface properties. The bulk wavenumber dispersion $q(\omega)$ and surface shear-stress $S_0$ from step (1), for a specified lateral surface displacement $u_0 e^{-i\omega t}$ is given by  
\begin{eqnarray}
q = &q_0 \Omega_0 R, & S_0 = s_0 u_0 \frac{\Omega_0}{R} tan\left(\Omega_0 R \Delta \right), \nonumber\\
R=&\sqrt{\frac{\Omega_0+\Omega_1+i}{\Omega_0 - i \Omega_1}}&
\end{eqnarray}
with $q_0 = \eta_0 /\sqrt{\rho G}$, $s_0 = G/q_0$, $\Omega_x = \omega \eta_x / G$, and $\Delta = d \sqrt{\rho G} / \eta_0$. 
The physical constants used are 
\begin{center}
\begin{tabular}{l l l}
$\rho$ & density & 1600 kg/m$^3$ \\
$G$ & shear modulus & 15.8 MPa (fit) \\
$\eta_0$ & series viscosity & 72.6 MPoise (fit) \\
$\eta_1$ & parallel viscosity & 54.9 MPoise (fit) \\
$\omega$ & frequency & $2 \pi$ 16565 rad/s \\
$d$ & layer thickness & 0-10 mm \\
\end{tabular}
\end{center}
with $G$ and $\eta_x$ determined by fitting to the data, as summarized in Fig.~\ref{fig3}. The skin depth is defined as the inverse of the imaginary part of the wavenumber, $\delta = 1/Im(q)$.

The relation to equivalent surface loading can then be found by equating the surface shear stress to that is required to move an rigid surface mass density, $\sigma$, attached to the surface, along with an effective viscous surface drag coefficient, $b$ defined a $F_{drag} = -b v_u^2$ acting on the surface. In the frequency domain, $(\sigma \omega^2 + i \omega b) u_0 = S_0$ so that 
\begin{equation}
\sigma = \sigma_0 \frac{1}{\Omega_0} {\cal R}e(\frac{1}{R}tan(\Omega_0 R \Delta),\,\, b = b_0 {\cal I}m(\frac{1}{R}tan(\Omega_0 R \Delta))
\label{eq:end}
\end{equation}
with $s_0 = \eta \sqrt{\rho /G}$ and $b_0 = \sqrt{\rho G}$.

Finally, the application of these relations to a free pillar resonant geometry is carried out by solving the 1D wave equation for a pillar torsion mode including the end loading forces. For now, the square pillar is analyzed assuming solid body rotation of pillar elements at each position along the torsion axis of the pillar, $\theta(z)=a e^{i q z} +b e^{-i q z}$, representing the forward and backward torsion waves in the pillar. The equations of motion are 
\begin{equation} 
\begin{pmatrix}
\sigma \omega^2  + i \omega b & -i q G e^{i q L} & i q G e^{-i q L} \\
-1 & e^{i q L} & e^{-i q L} \\
0  & 1 & -1 \\
\end{pmatrix}
\begin{pmatrix}
\theta_0 \\
a \\
b \\
\end{pmatrix}=0
\label{eq:pillar}
\end{equation}
where $\theta_0$ is the displacement at the end-loaded surface, $0<z<L$. Note that here, $\rho_{Al}$ and $G_{Al}$ are the respective density and shear modulus of the aluminum pillar. The first row expresses the displacement at the top ($z=L$), the second matches the end torsion strain with the surface loading quantities of Eq.~(\ref{eq:end}) at the top, and the third expresses the free end at the bottom, $z=0$. 

In the weakly-loaded limit, the solutions show that the the lowest mode frequency shift, $\Delta \omega$, is related to the loading parameters by
\begin{equation}
\sigma = -\rho L \frac{Re(\Delta \omega) }{\Omega_0}, \hspace{1cm} b = -\pi \sqrt{\rho G}   \frac{Im(\Delta \omega) }{\omega_0}.
\label{eq:end_result}
\end{equation}
These are the final relations necessary to identify the putty material properties with the pillar frequency shifts shown in Fig.~\ref{fig3}.

{\bf Fitting Relations for Extraction the Eigenfrequencies:} The measurement procedure for the experimental transmission spectrum is as follows: clamp one pillar; add a putty ball to the other, measuring its transmission spectra with the drive and detector set on that pillar; remove the clamp; switch the drive set to that of the released pillar, measuring the two-pillar transmission spectrum. This process is repeated until the exceptional 
point was surpassed. 

To extract information of the position of both resonances as precisely as possible (the real part of the frequency) as well as their linewidths (the imaginary part of the frequency), we consider the complex frequency domain ($e^{i \omega t}$) relations for an ideal pair of coupled simple harmonic oscillators, individually damped, excited by a force acting on one oscillator, and detected on the other oscillator, defined earlier. This time, the individual drag constants in $F_{drag} = -bv$ are $b_d$, for the driven side, and $b_{\Gamma}$ for the detected side, and amplitude $F_d$ on the driven side. Experimentally, the mass under the action of the force has the inherent damping value $\gamma_d$ while the detected mass receives the incremental damping, $\Gamma$. The resulting particle displacements for this model are given in terms of scaled parameters by
\begin{equation}
\begin{pmatrix}
x_d \\
x_{\Gamma}
\end{pmatrix}
=
\begin{pmatrix}
1 + \kappa + iu\gamma_d - u^2 & -\kappa \\
-\kappa & 1 + \kappa + \epsilon + iu\Gamma - u^2
\end{pmatrix}
\begin{pmatrix}
f_d \\
0
\end{pmatrix}
\label{eq:lossloss}
\end{equation}
where $x_d$ and $x_{\Gamma}$ are the displacements of the driven and detected masses, and $u=\omega/\omega_0$, $f_d=F_d/k_0$, and $\kappa = k/k_0$. The additional parameter $\epsilon$ is a detuning parameter applied to the natural frequency of the damped (un-driven) oscillator included as a check on the mass loading associated with the experimentally applied damping mechanism. The complex transmittance expression, Eq.~\ref{transmit} used to fit the data is the displacement $x_{\Gamma}(u)$ with added factors $\gamma_d$ and $A$ in the numerator to re-define the transmission amplitude $A$ as the resonant contribution, and $t_0$ and $\phi$ accounting for an overall transmission level and phase shift introduced by the impedance transformers and transducer mechanical mismatch.


\end{document}